**Potential order-of-magnitude enhancement of wind farm power density via counter-rotating vertical-axis wind turbine arrays**

John O. Dabiri

Graduate Aeronautical Laboratories & Bioengineering, California Institute of Technology

Pasadena, California 91125, USA

Correspondence:

Mail Code 138-78

1200 E. California Blvd.

Pasadena, CA 91125

Phone: 1-626-395-6294

Fax: 1-626-577-5258

Email: jodabiri@caltech.edu

**Abstract**

Modern wind farms comprised of horizontal-axis wind turbines (HAWTs) require significant land resources to separate each wind turbine from the adjacent turbine wakes. This aerodynamic constraint limits the amount of power that can be extracted from a given wind farm footprint. The resulting inefficiency of HAWT farms is currently compensated by using taller wind turbines to access greater wind resources at high altitudes, but this solution comes at the expense of higher engineering costs and greater visual, acoustic, radar and environmental impacts. We investigated the use of counter-rotating vertical-axis wind turbines (VAWTs) in order to achieve higher power output per unit land area than existing wind farms consisting of HAWTs. Full-scale field tests of 10-m tall VAWTs in various counter-rotating configurations were conducted under natural wind conditions during summer 2010. Whereas modern wind farms consisting of HAWTs produce 2 to 3 watts of power per square meter of land area, these field tests indicate that power densities an order of magnitude greater can potentially be achieved by arranging VAWTs in layouts that enable them to extract energy from adjacent wakes and from above the wind farm. Moreover, this improved performance does not require higher individual wind turbine efficiency, only closer wind turbine spacing and a sufficient vertical flux of turbulence kinetic energy from the atmospheric surface layer. The results suggest an alternative approach to wind farming that has the potential to concurrently reduce the cost, size, and environmental impacts of wind farms.

**Introduction**

A principal challenge for all forms of renewable energy is that their sources—solar radiation or wind, for example—are more diffuse than fossil fuels. As a consequence, existing renewable energy technologies require substantial land resources in order to extract appreciable quantities of energy. This limitation of land use is especially acute in the case of wind energy, which currently faces an additional constraint in that conventional propeller-style wind turbines (i.e. horizontal-axis wind turbines; henceforth, HAWTs) must be spaced far apart in order to avoid aerodynamic interference caused by interactions with the wakes of adjacent turbines. This requirement has forced wind energy systems away from high energy demand population centers and toward remote locations including, more recently, offshore sites. It has also necessitated the implementation of very large wind turbines, so that the inefficiency of the wind farm as a whole can be compensated by accessing the greater wind resources available at high altitudes. However, this solution comes at the expense of higher engineering costs and greater visual, acoustic, radar and environmental impacts. These issues represent a principal barrier to the realization of wind energy technology that is both economically viable and socially acceptable (1, 2).

To maintain 90 percent of the performance of isolated HAWTs, the turbines in a HAWT farm must be spaced 3 to 5 turbine diameters apart in the cross-wind direction and 6 to 10 diameters apart in the downwind direction (1, 2). The power density of such wind farms, defined as the power extracted per unit land area, is between 2 and 3 W m$^{-2}$ (3).

Wind turbines whose airfoil blades rotate around a vertical axis (i.e. vertical-axis wind turbines; henceforth, VAWTs) have the potential to achieve higher power densities than HAWTs. This possibility arises in part because the swept area of a VAWT rotor (i.e. the cross-sectional area that interacts with the wind) need not be equally apportioned between its breadth—which determines the size of its footprint—and its height. By contrast, the circular sweep of HAWT blades dictates that the breadth and height of the rotor swept area are identical. Therefore, whereas increasing HAWT rotor swept area necessarily increases the turbine footprint, it is possible to increase the swept area of a VAWT independent of its footprint, by increasing the rotor blade height. Table 1 compares the power density of a commercially-available VAWT with two common HAWT models. The power density of the VAWT design is more than three times that of the HAWTs, suggesting that VAWTs may be a more effective starting point than HAWTs for the design of wind farms with high power density.

The turbine power densities indicated in Table 1 are not achieved in practice due to the aforementioned spacing requirements between turbines in a wind farm. However, we hypothesized that counter-rotating arrangements of VAWTs can benefit from constructive aerodynamic interactions between adjacent turbines, thereby mitigating reductions in the performance of the turbines when in close proximity. By accommodating a larger number of VAWTs within a given wind farm footprint, the power density of the wind farm is increased. Furthermore, by capturing a greater proportion of the wind energy incident on the wind farm footprint, it becomes unnecessary to use wind turbines as large as those commonly found in modern HAWT farms. In turn, the use of smaller turbines can reduce the complexity and cost of the individual wind turbines, since the smaller wind turbines do not experience the high gravitational, centrifugal, and wind loading that must be withstood by large HAWTs. The less severe design requirements can enable implementation of less expensive materials and manufacturing processes.

Here we present an initial study of this concept of counter-rotating VAWT farms, by measuring wind turbine performance at full scale and in naturally-occurring wind conditions. Although field measurements lack the controllable environment of scale model experiments in a wind tunnel or numerical simulations, they do provide the most direct support of the validity of the proposed wind farm concept. The data set presented here can also be used as a baseline for comparison with future scale model experiments and numerical simulations.

**Materials and methods**

*Field site summary*

Experiments were conducted at a field site in the Antelope Valley of northern Los Angeles County, California, USA. The site is vacant desert and the topography is flat for approximately 1.5 kilometers in all directions (figure 1*A*). Over the duration of these experiments, from June to September 2010, the mean wind speed was approximately 7.8 m/s at 10 m with mean turbulence fluctuations (i.e. standard deviation) of 2.6 m/s. Figure 2 plots the daily average wind speed and turbulence fluctuations during the course of the experiments. Figure 3 plots the distribution of wind direction at the site; the prevailing wind is from the southwest. The natural variability of the wind direction enabled the sensitivity of turbine performance to wind direction to be studied without requiring a large number of discrete turbine configurations to be tested (see VAWT positioning and protocols below).

*Wind turbine design*

The field tests utilized six 10-m tall x 1.2-m diameter VAWTs. The turbines were a modified version of a commercially available model (Windspire Energy Inc.) with 4.1-m span airfoil blades and a 1200-W generator connected to the base of the turbine shaft. Three of the turbines rotated around their central shaft in a clockwise direction (e.g. from a top view) in winds above 3.8 m s$^{-1}$; the other three rotated in a counter-clockwise direction when the wind speed exceeded the same threshold (henceforth, the cut-in wind speed) .

*VAWT positioning and protocols*

Each of the experiments was conducted with the turbines positioned within the same 75 m x 75 m tract of land. One of the six turbines remained fixed in the same location for all of the experiments. The remaining turbines were manually repositioned on portable footings in order to create the various configurations studied. The schedule of turbine positions is listed in Table 2, along with the number of hours that each turbine configuration was measured.

*Turbine measurements*

The rotational speed and electrical power generated by each turbine were monitored in real-time and recorded at 1 Hz using custom software designed to interface with the turbines (WindSync, Windspire Energy Inc.). Measurement accuracy was ± 5 percent for both parameters. Each measurement was assigned a time stamp that was synchronized with separately collected meteorological data (see Meteorological measurements section below) and was manually uploaded via a satellite uplink (HughesNet) from the field site to a computer at the California Institute of Technology, where the data was analyzed.

*Meteorological measurements*

A 10-m meteorological tower was erected at the northwest corner of the field site in order to measure wind speed and direction at a height comparable to the mid-span height of the VAWT blades (8 m). The tower was located 15 turbine diameters northwest (i.e. approximately cross-wind) of the nearest VAWT to ensure that it did not affect the wind conditions near the turbines. Although the need to avoid aerodynamic interference between the meteorological tower and the VAWTs precluded wind measurements using the tower closer to the turbines, the difference in their position was significantly smaller than the length scale over which mean flow in the atmospheric surface layer changes (4, 5). To be sure, the turbulence fluctuations, which were typically 30 to 40 percent of the mean wind speed, likely overwhelm differences between the instantaneous wind speed at the location of the meteorological tower and at the turbines.

The accuracy of the wind speed sensor (Thies First Class) and wind direction sensor (Met One) measurements was ± 3 percent and ± 5 degrees, respectively. Data from the meteorological tower was recorded at 1 Hz using a datalogger (Campbell Scientific). The data was assigned a timestamp synchronized with the turbine measurement data before transmission via the satellite uplink.

*Power coefficient calculation*

The turbine power coefficient is defined as the fraction of incident kinetic energy passing through the swept area of the turbine rotor that is converted to electrical energy (2). In terms of the generated electrical power $P$, air density $\rho$, turbine rotor swept area $A$ (equal to the product of the turbine rotor diameter and height), and wind speed $U$, the power coefficient is

$$C_p = \frac{P}{(1/2)\rho A U^3},$$ (1)

where the air density was estimated to be 1.2 kg m$^{-3}$ and the turbine rotor swept area is 5.02 m$^2$.

*Wind farm power density calculation*

The wind farm power density is defined as the electrical power generated by the wind farm divided by the area of its footprint (3). In terms of the turbine rated power $P$, capacity factor $C$, wind farm aerodynamic loss factor $L$, wind turbine spacing $S$ and wind turbine diameter $D$, the wind farm power density is

$$WPD = \frac{PC(1-L)}{(\pi/4)(SD)^2},$$ (2)

where the factor $\pi/4$ arises due to the assumption that each turbine has a circular footprint with diameter ($S$ x $D$) inside which no other turbines can be located.

**Results**

In the first set of experiments, we measured the performance of two counter-rotating VAWTs whose axes of rotation were separated by 1.65 turbine diameters (Figure 1*B*). The clockwise-rotating turbine (denoted *CW1*) was measured at multiple positions around the azimuth of the counter-clockwise-rotating turbine (denoted *CCW1*) in order to determine the dependence of turbine performance on the relative direction of the incident wind. In addition, the performance of turbine *CCW1* was measured while it was isolated (i.e. separated by 10 turbine diameters from turbine *CW1*), in order to evaluate the effect of the close proximity of the turbines on the power coefficient (i.e. the fraction incident wind energy that is converted to electrical energy, denoted $C_p$). A normalized power coefficient, $C_p^{norm}$, defined as the ratio of the turbine power coefficient in the counter-rotating configuration to the power coefficient of the isolated turbine, was used to evaluate the performance of each configuration.

The normalized power coefficient of turbine *CCW1* (and, by spatial symmetry, the normalized power coefficient of turbine *CW1*) was nearly insensitive to the incident wind direction over the 315 degrees of wind direction variation that was observed (Figure 4*A*). Averaged over all incident wind directions, the close proximity of the turbines slightly improved their performance relative to the turbines in isolation (Figure 4*B*). This is in contrast to typical performance reductions between 20 and 50 percent for HAWTs at a similar turbine spacing (6-9). The result is qualitatively consistent with the predictions of previous simplified numerical models, which anticipated that closely-spaced VAWTs can reciprocally enhance the wind field of the adjacent turbines (10, 11).

In a second set of experiments, we studied the performance of a third VAWT placed 1.65-diameters downwind from two counter-rotating VAWTs with the same spacing (Figure 1*C*). These experiments explored the effect of downwind blockage caused by the two closely-spaced upwind turbines. We observed a significant decrease in the performance of the downwind turbine, especially at higher ratios of rotor blade tip speed to wind speed (henceforth, tip speed ratio, 12). However, when the spacing of the downwind turbine was increased to four diameters, its performance was recovered to within 5 percent of the isolated turbine performance across the range of observed tip speed ratios (Figure 5). This rapid recovery of the downwind flow field is in marked contrast to the 15 to 20 diameters of downwind spacing found to be required for a similar level of wake recovery in a recent numerical simulation of a large HAWT (13).

Based on the preceding experiments, we hypothesized that by increasing the mean spacing of all turbines in an array to four diameters, upstream blockage effects would be significantly reduced. Figure 1*D* illustrates the wind farm configuration implemented in field tests. Nearest-neighbor turbines were counter-rotating in order to take advantage of the lesser aerodynamic interference between counter-rotating VAWTs as compared to co-rotating VAWTs (10, 11). The field tests confirmed that each of the downwind turbines in the array achieved performance comparable to the VAWT at the front of the array (Figure 6*A*). The performance of the turbine located five positions downwind from the front of the array was reduced by less than five percent relative to the farthest upwind turbine, which is within the measurement uncertainty.

Averaged over the 48.6-$m^2$ footprint of the six-turbine VAWT array, the daily mean power density produced by the array varied from 21 to 47 W $m^{-2}$ at wind speeds above cut-in and 6 to 30 W $m^{-2}$ overall (Figure 6*B*). This performance significantly exceeded the 2 to 3 W $m^{-2}$ power density of modern HAWT farms, despite the relatively low mean wind speed during this set of field tests (5.7 m $s^{-1}$).

To be sure, practical limitations on the number of VAWTs in the field tests precluded a direct evaluation of turbines surrounded on all sides by neighboring VAWTs, as would be the case for the majority of turbines in a wind farm. To extrapolate the present measurements to larger VAWT farms, we considered the present VAWT diameter (1.2 m) and inter-turbine spacing (4 diameters), and we made conservative estimates for both the total aerodynamic loss in the array (10 percent) and the capacity factor (i.e. the ratio of actual power output to the maximum generator power output; 30 percent). The calculated power density for a VAWT farm with these parameters is approximately 18 W m$^{-2}$ (cf. equation 2). This performance is 6 to 9 times the power density of modern wind farms that utilize HAWTs (14).

Furthermore, it is straightforward to compute combinations of VAWT rated power output and turbine spacing that can achieve 30 W m$^{-2}$ (i.e. 10 times modern HAWT farms) by using 1.2-m diameter VAWTs like those studied here (Figure 7). Higher VAWT rated power outputs can be achieved by taller turbine rotors than the 4.1-m structures used in these experiments, and by connecting the turbine shaft to larger generators. Indeed, in initial field tests with 6.1-m tall rotors, the captured wind power exceeded the capacity of the 1200 W generator on each turbine.

**Discussion**

The large increases in wind farm power density demonstrated here may be surprising when one considers that the efficiency (i.e. power coefficient) of modern HAWTs approaches the theoretical upper limit of 59.2 percent aerodynamic efficiency for isolated HAWTs (2). The present results suggest that the physical limit on wind energy extraction using the VAWT array approach is not the individual turbine efficiency, as is the case for well-spaced HAWTs that essentially operate in isolation within a wind farm. Instead, wind energy extraction is limited by the wind resource itself, especially the horizontal wind speed and the vertical flux of turbulence kinetic energy required to transport wind energy to turbines downwind from the front of the wind farm. This upper limit, which is based on properties of the atmospheric surface layer and the surface roughness created by the wind turbines themselves (4, 5, 15, 16), supersedes the theoretical limit on isolated HAWT efficiency as the primary determinant of maximum VAWT farm performance. Stated differently, although individual VAWTs often exhibit lower power coefficients than HAWTs (2), this deficiency is compensated (indeed, overcompensated) by the fact that VAWTs can be placed closer together. The wind energy that is not extracted by one VAWT (due to its inefficiency) can be collected by an adjacent VAWT in close proximity.

To quantify the upper limit on wind energy extraction from VAWT arrays, we considered the horizontal (i.e. from upwind) and vertical (i.e. from above) fluxes of kinetic energy into the wind farm. These power sources, denoted $P_{horz}$ and $P_{vert}$, respectively, can be estimated as (5, 15)

$$P_{horz} \approx \tfrac{1}{2}\rho A_{frontal} U^3 \tag{3}$$

$$P_{vert} \approx -\rho A_{planform} U \langle u'w' \rangle , \tag{4}$$

where $\rho$ is the air density, $U$ is the mean horizontal wind speed, $u'$ is the horizontal turbulence velocity fluctuation, $w'$ is the vertical turbulence velocity fluctuation, $A$ is the frontal or planform (i.e. top view) area, respectively, and the angle brackets denote an ensemble average.

The Reynolds stress $\langle u'w' \rangle$ can be estimated in terms of the friction velocity $u_*$ as (5)

$$-\langle u'w'\rangle = u_*^2 = \left[\frac{U\kappa}{\ln\left((z-d)/z_0\right)}\right]^2, \tag{5}$$

where $\kappa$ is von Karman's constant $\approx 0.4$, $z$ is the height above the ground, and $d$ and $z_0$ are, respectively, the zero plane displacement (i.e. the effective height at which the surface roughness acts) and roughness length of the VAWT array. Per convention, the values of $d$ and $z_0$ are taken as 2/3 and 1/10 of the turbine height, respectively (5).

For the present experiments, wherein $\rho = 1.2$ kg m$^{-3}$ and $U = 7.8$ m s$^{-1}$ at 10 m above the ground (averaged over all field tests, see Materials and methods), the input flux of kinetic energy from upwind is approximately 285 W per square meter of frontal area. This frontal kinetic energy flux will limit the performance of VAWTs near the front of the array; however, the majority of the turbines in a large VAWT farm will be limited by the lower planform kinetic energy flux from above the wind farm (15, 16). Figure 6*B* indicates that the wind farm power density is correlated with, and indeed bounded by, the planform kinetic energy flux. Above the wind farm, the mean wind speed will be reduced from its upwind value due to the elevated surface friction caused by the presence of the wind turbines. Figure 8 plots the planform kinetic energy flux model from equations (4) and (5) as a function of the ratio of the reduced mean wind speed $U_r$ to the unperturbed wind speed $U$ (i.e. in the absence of the wind farm). For comparison, the nominal performance of modern HAWT farms is also shown. The results suggest that as long as the wind speed above the wind farm remains greater than 1/3 of the unperturbed wind, the VAWT farm performance upper bound dictated by the planform kinetic energy flux exceeds the performance of current HAWT farms. For $U_r/U > 0.75$, the VAWT farm planform kinetic flux is an order of magnitude greater than the performance of modern HAWT farms.

The present measurements are insufficient to determine the range of $U_r/U$ that can be achieved in practice for large-scale VAWT farms. The value will depend the local stability of the atmospheric surface layer, the spatial density and height profile of the VAWTs, and their effective drag properties. Further study of the interplay among these parameters is essential and is a focus of ongoing and future research.

By including periodic gaps of larger downwind spacing and/or turbine height variations between clusters of downwind VAWTs, it may also be possible to prevent saturation of the frontal kinetic energy flux without significantly compromising the gains in wind farm power density. With regard to the former strategy of downwind spacing, we verified that by removing the turbine immediately upwind of the rearmost VAWT in the present array, its performance was further improved (Figure 6$A$, red dash-dot curve).

Counter-rotation of adjacent VAWTs is important because it ensures that the airflow induced by each of the turbines in the region between them is oriented in the same direction (17, 18; see also Figure 9). Hence, the creation of horizontal wind shear (i.e. velocity gradients), which leads to turbulence and energy dissipation in the region between the turbines, is reduced relative to adjacent turbines that rotate in the same direction (19, 20). Since the remaining wind energy between turbines is not dissipated by turbulence, it can be subsequently extracted by VAWTs located further downwind. This process is most effective for VAWTs operating at higher tip speed ratios (i.e. greater than 2), since in this regime the turbine rotation can suppress vortex shedding and turbulence in the wake in a manner similar to that observed in previous studies of spinning cylinders (21-23). At lower tip speed ratios, the VAWTs likely create a larger wake akin to that of a stationary cylinder; we observed correspondingly reduced performance in the present field tests.

The overall approach described presently is fundamentally different from current practices in wind energy harvesting: here, a large number of smaller VAWTs are implemented instead of fewer, large HAWTs. The higher levels of turbulence near the ground—both naturally occurring and induced by the VAWT configuration—enhance the vertical flux of kinetic energy delivered to the turbines, thereby facilitating their close spacing. This approach has the potential to concurrently alleviate many of the practical challenges associated with large HAWTs, such as the cost and logistics of their manufacture, transportation and installation (e.g. by using less expensive materials and manufacturing processes, and by exploiting greater opportunities for mass production); environmental impacts (e.g. bird and bat strikes); acoustic and radar signatures (e.g. lower tip speed ratios than HAWTs, 2); visual signature (Figure 10); and general acceptance by local communities. These issues, although not strictly scientific, limit the further expansion of existing wind energy technology.

The present results encourage a search for optimal configurations of counter-rotating VAWTs that can improve upon the power density achieved here. Such optimal solutions may achieve enhanced turbine performance in close proximity (e.g. Figure 4) while minimizing downwind blockage effects and enhancing the vertical flux of kinetic energy via manipulation of the zero plane displacement and roughness length of the VAWT array. Finally, we note that the energy harvesting principles developed here are equally applicable to underwater turbines in the ocean.

**Acknowledgments**

The author gratefully acknowledges funding from the National Science Foundation Energy for Sustainability program (CBET-0725164) and the Gordon and Betty Moore Foundation. The author also thanks R. W. Whittlesey for providing assistance in establishing the satellite data connection to the field site.

19. To be sure, the second-nearest-neighbor turbines in the array are co-rotating. However, their separation is a factor $2^{1/2}$ greater than the separation of nearest-neighbor counter-rotating turbines. Since the effect of induced flow decays with radial distance as $r^{-2}$ or faster (18), the magnitude of the co-rotating interactions is at most 1/2 of the interactions between counter-rotating turbines.

**Table 1.** Comparison of VAWT and HAWT power density. The power density is calculated as the turbine rated power divided by the area of the circular footprint swept by the turbine rotor blades when rotated in yaw by 360 degrees.

| Turbine Type | Rated Power (MW) | Rotor Diameter (m) | Power Density (W/m$^2$) |
|:---:|:---:|:---:|:---:|
| VAWT | 0.0012 | 1.2 | 1061 |
| HAWT | 2.5 | 100 | 318 |
| HAWT | 3.0 | 112 | 304 |

**Table 2.** Field test schedule. See text and Figure 1 for definitions of abbreviations.

| Test Dates | Turbine Configuration | Measurement Duration (continuous) |
|---|---|---|
| 12 JUN - 23 JUN | *CW1* south of *CCW1*, 1.65-dia. separation | 252 hours |
| 25 JUN - 7 JUL | *CW1* north of *CCW1*, 1.65-dia. separation | 312 hours |
| 9 JUL - 23 JUL | *CW1* south of *CCW1*, 10-dia. separation | 360 hours |
| 30 JUL - 11 AUG | *CW1* west of *CCW1*, 1.65-dia. separation | 312 hours |
| 13 AUG - 15 AUG | *CW2* south of *CCW2*, 1.65-dia. separation<br>*CW3* northeast of *CCW2*, 1.65-dia. separation | 72 hours |
| 13 AUG - 17 AUG | *CW1* east of *CCW1*, 1.65-dia. separation,<br>*CW1* rotor stationary | 120 hours |
| 19 AUG - 29 AUG | *CW2* south of *CCW2*, 1.65-dia. separation<br>*CW3* northeast of *CCW2*, 4-dia. separation | 264 hours |
| 30 AUG - 1 SEP | *CW3* northwest of *CCW2*, 14-dia. separation | 58 hours |
| 3 SEP - 5 SEP | Fig. 1*D*, last downwind *CW* turbine absent | 48 hours |
| 10 SEP - SEP 20[†] | Fig. 1*D* | 251 hours |

† *CW* turbine in right column of Fig. 1*D* measured 10-11 SEP and 18-20 SEP only. *CCW* turbine in middle column of Fig. 1*D* measured 10-13 SEP only.

**Figure Legends**

**Figure 1.** Vertical-axis wind turbine (VAWT) configurations. (*A*) View of field site toward southwest (approximately upwind). Each turbine is 10 m tall to the top of the rotor blades. Three-turbine array is at left, two-turbine array is in center. Inset at right indicates height of the turbines relative to a 1.9-m tall person. (*B*) Schematic top view of two-VAWT configurations. Top of panel is due north. Circles indicate 1.2-m turbine diameter, arrows indicate direction of turbine rotation. Turbine spacing (i.e. 1.65 turbine diameters) is indicated by the length of the single grey lines and is drawn to scale. Red circle, turbine *CCW1*; blue circle, turbine *CW1*; black circles, additional positions of turbine *CW1* tested during measurements of wind direction sensitivity. Black arrow at lower left indicates prevailing wind direction in panels *B-D* (see Figures 2 and 3 for full distributions of wind speed and direction, respectively). (*C*) Schematic top view of three-VAWT configurations. Blue circles (i.e. clockwise-rotating turbines) are spaced 1.65 turbine diameters from red turbine (i.e. counter-clockwise-rotating turbine), as indicated by the length of the single grey lines.  Black circle, alternate position of upper blue circle at 4 turbine diameters downwind, as indicated by the length of the double grey lines. (*D*) Schematic top view of six-VAWT configuration. Red and blue circles indicate positions of six VAWTs during measurements. Length of double grey lines indicates 4 turbine diameter spacing. Grey circles indicate additional turbine positions in a hypothetical larger-scale array.

**Figure 2.** Measured daily average wind speed (solid line) and standard deviation turbulence fluctuations (dashed band) over the duration of field tests.

**Figure 3.** Histogram of measured wind direction. Angle coordinate is measured in degrees from north. Radial coordinate is the number of hours observed for each wind direction.

**Figure 4.** Measurement of two-VAWT configuration with 1.65 turbine diameter separation (see Fig. 1$B$). ($A$) Plot of normalized power coefficient $C_p^{norm}$ (radial coordinate) versus incident wind direction (angle coordinate in degrees from north). Inset turbine schematic indicates position of VAWTs relative to incident wind. Length of grey line indicates 1.65 turbine diameter spacing. Wind directions observed for less than 900 s are omitted (i.e. incident wind from the north). Values of $C_p^{norm} = 1$ indicate turbine performance equal to that of the isolated turbine. ($B$) Solid line, plot of normalized power coefficient $C_p^{norm}$ versus tip speed ratio for all incident wind directions. The tip speed ratio is given by $(\pi D \Omega) U^{-1}$, where $D$ is the wind turbine rotor diameter, $\Omega$ is the turbine rotation rate, and $U$ is the wind speed. Vertical dotted line indicates designed operating tip speed ratio of turbines.

**Figure 5.** Normalized power coefficient $C_p^{norm}$ of turbine *CW3* (upper clockwise turbines in Fig. 1$C$) versus turbine tip speed ratio. Prevailing wind direction is indicated by black arrow at lower left of Fig. 1$B$. Blue curve, 1.65-diameter downwind spacing from counter-rotating upwind turbine pair (i.e. upper blue circle in Fig. 1$C$); black curve, 4-diameter downwind spacing (i.e.

upper black circle in Fig. 1*C*). Values of $C_p^{norm}$ = 1 indicate turbine performance equal to that of the isolated turbine. Vertical dotted line indicates designed operating tip speed ratio of turbines.

**Figure 6.** Performance in counter-rotating six-VAWT configuration. (*A*) Plot of normalized power coefficient $C_p^{norm}$ versus tip speed ratio for all incident wind directions. Data are normalized by the power coefficient of the farthest upwind turbine (i.e. *CW* turbine in left column of Fig. 1*D*). Dotted red curve, *CCW* turbine in left column of Fig. 1*D*; dashed red curve, *CCW* turbine in middle column; solid red curve, *CCW* turbine in right column; dash-dot red curve, *CCW* turbine in right column with adjacent *CW* turbine removed; dashed blue curve, *CW* turbine in middle column; solid blue curve, *CW* turbine in right column. Vertical dotted line indicates designed operating tip speed ratio of turbines. (*B*) Measured array power density versus planform kinetic energy flux (see text for definition). Data points are labeled according to measurement date. Closed circles, 24-hour average (except 10 Sep, which is an average from 13:00 to 24:00); open circles, average above cut-in wind speed.

**Figure 7.** Turbine rated power and spacing combinations for order-of-magnitude increase in wind farm power density relative to existing HAWT farms. Blue curve, 30 W m$^{-2}$ wind farm power density. Curve assumes 1.2-m turbine diameter as in the present tests, 30 percent turbine capacity factor, and 10 percent power loss due to aerodynamic interactions within the VAWT array. Dashed grey curves correspond to the power densities of existing renewable energy technologies (3).

**Figure 8.** Planform kinetic energy flux versus the ratio of mean wind speed above the wind farm $U_r$ to the unperturbed mean wind speed $U$ (i.e. in the absence of the wind farm). The planform kinetic energy flux is correspondingly reduced with $U_r$ replacing $U$ in equations (4) and (5). For mean wind speeds that are greater than approximately 1/3 of the unperturbed wind speed, the planform kinetic energy flux exceeds the performance of current HAWT farms (black dashed line). For $U_r/U > 0.75$, the VAWT farm planform kinetic flux is an order of magnitude greater than the performance of modern HAWT farms.

**Figure 9.** Schematic of induced airflow between co-rotating VAWTs (panel $A$) and counter-rotating VAWTs (panel $B$). Co-rotating VAWTs (circles) induce airflow (hollow arrows) in opposite directions, whereas counter-rotating VAWTs (circles) induce airflow (hollow arrows) in the same direction.

**Figure 10.** Visual signature of VAWT array. Image taken approximately 1 km from test facility (indicated by white arrow). 10 m height of VAWTs is labeled at right, in addition to approximate 100 m height of a typical large HAWT. Photo credit: R. W. Whittlesey.

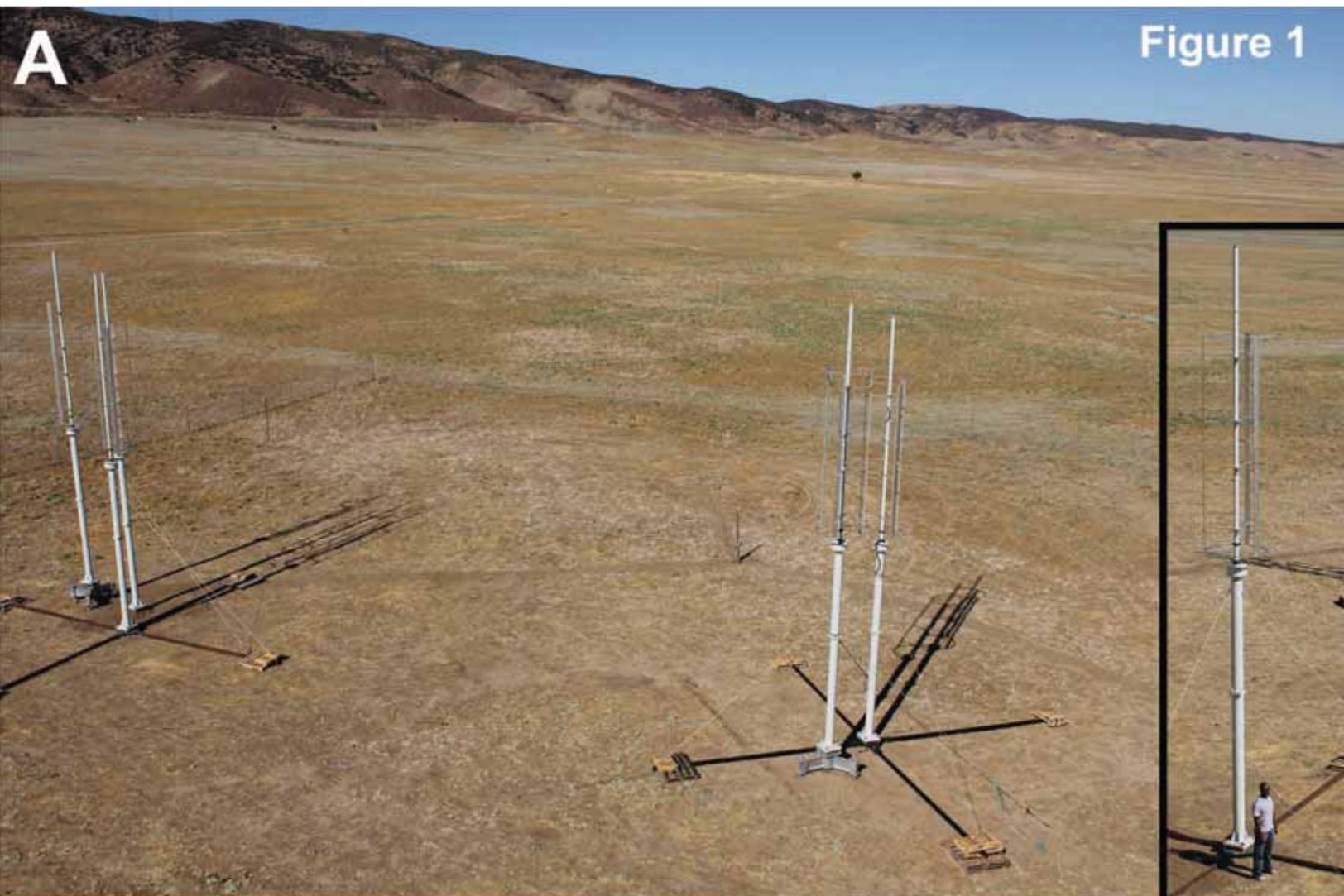

**Figure 1**

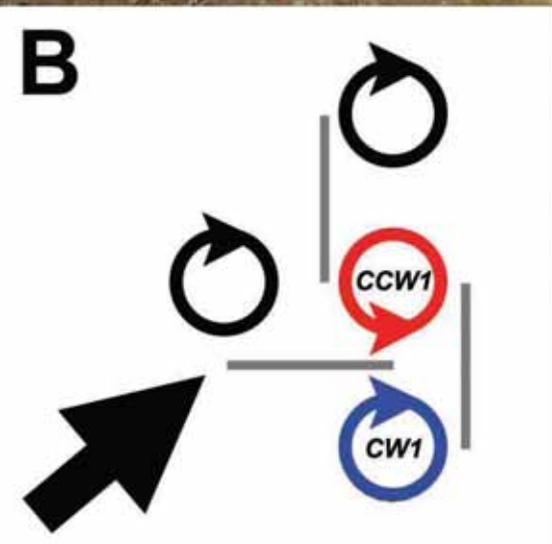

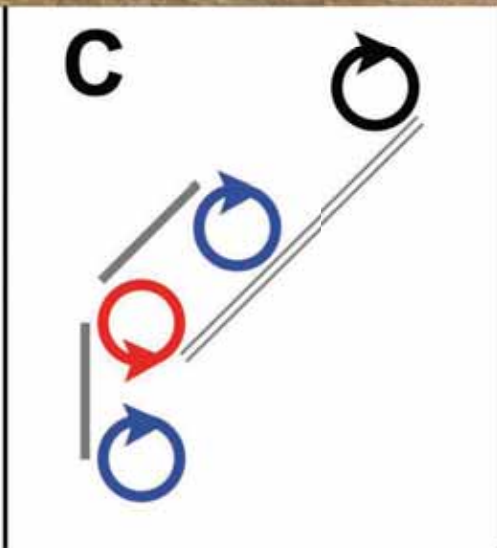

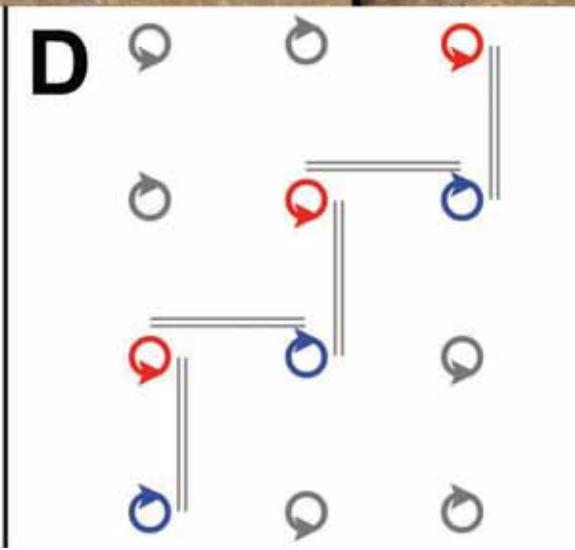

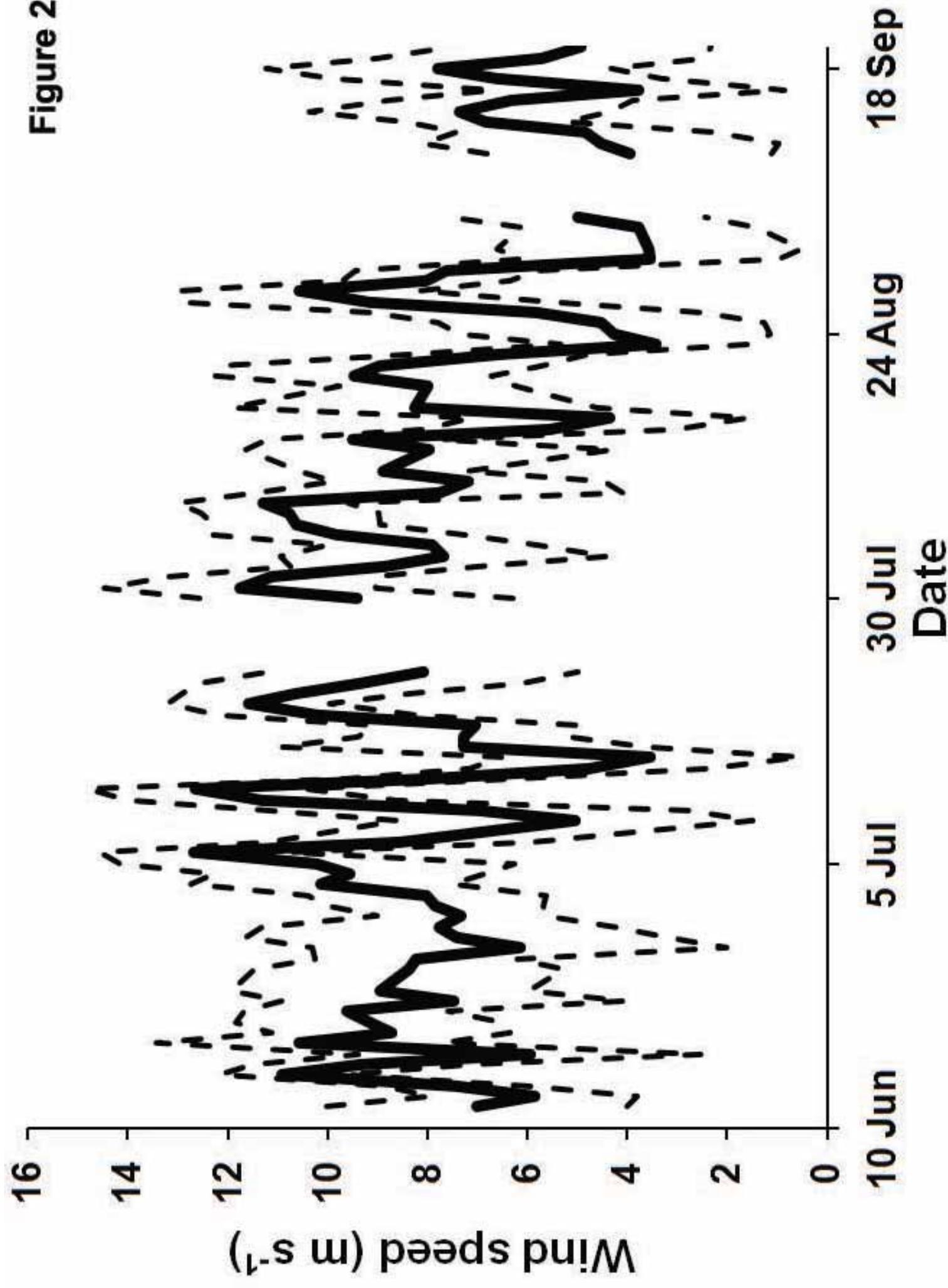

Figure 2

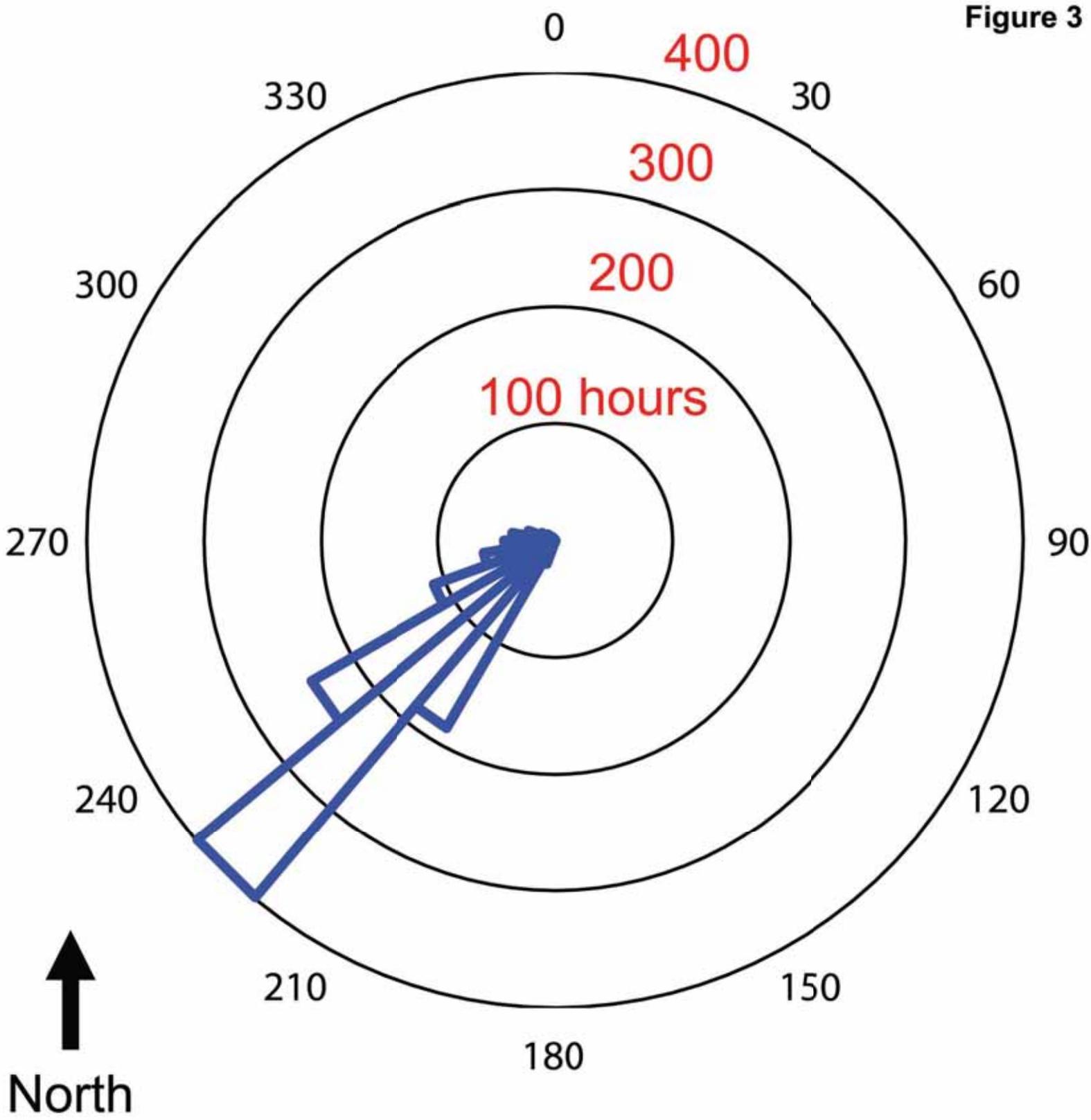

**Figure 3**

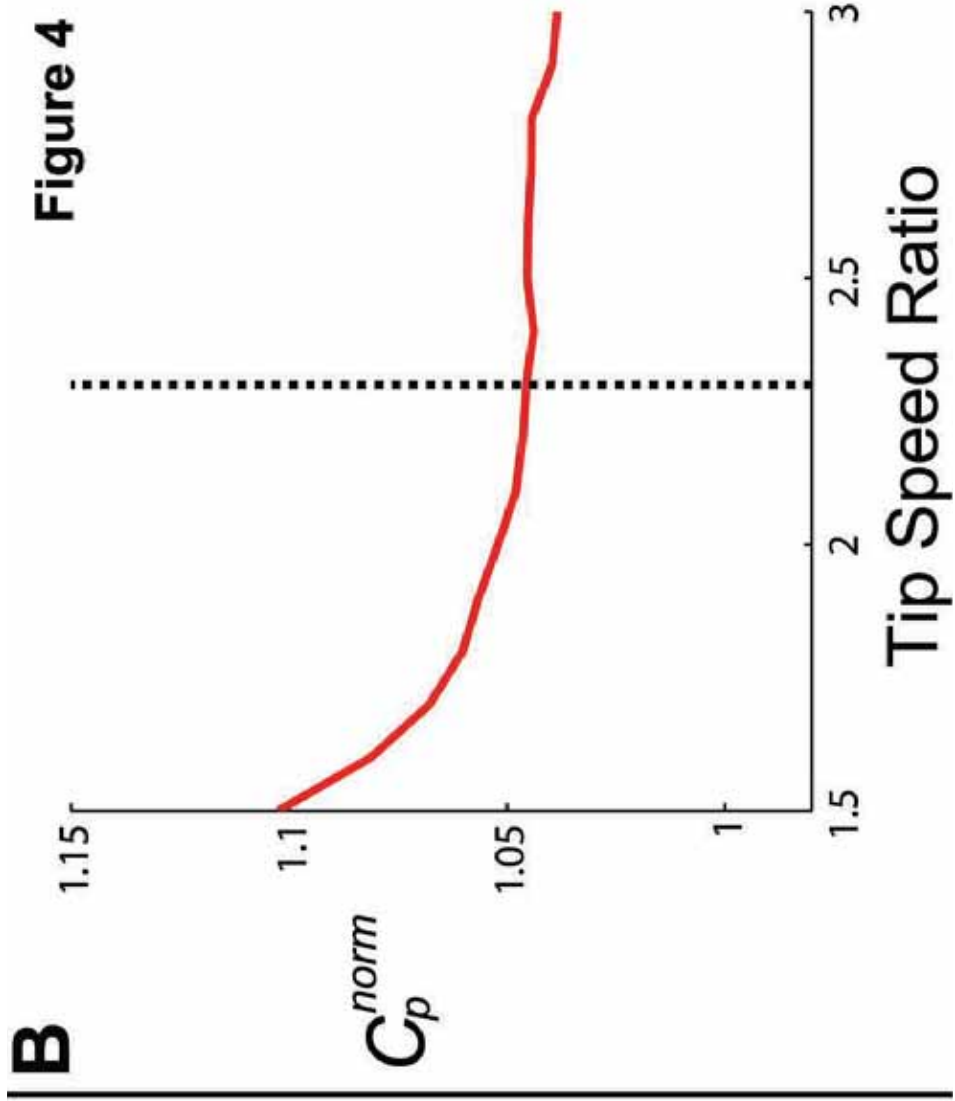

**Figure 4**

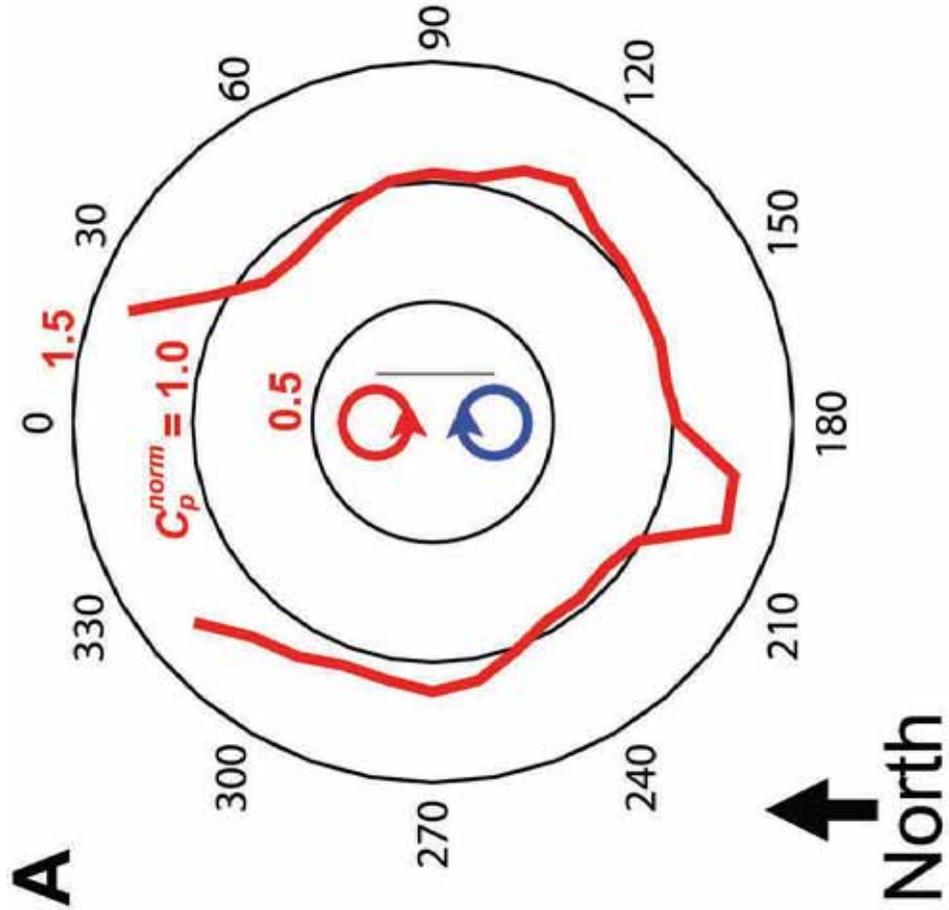

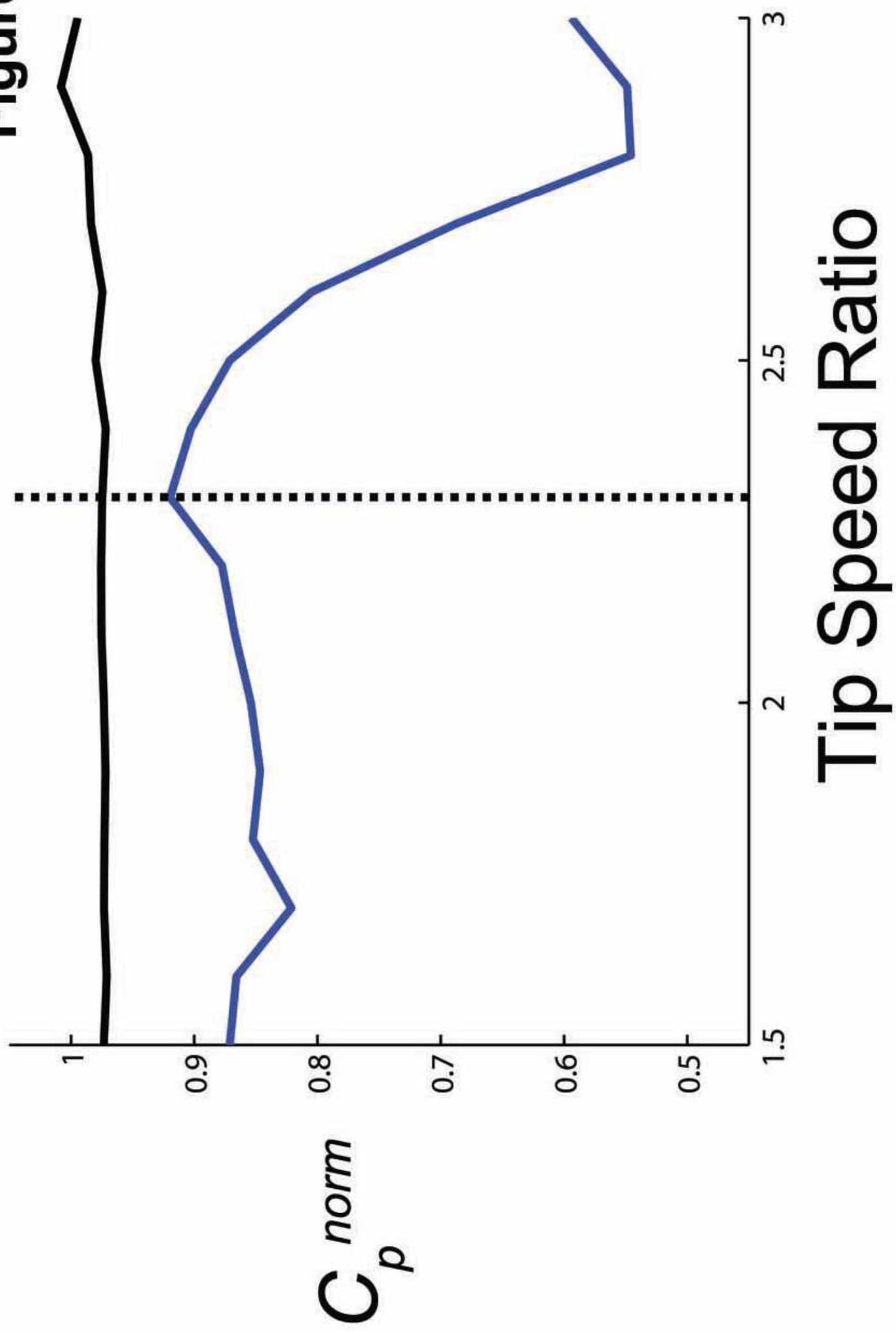

Figure 5

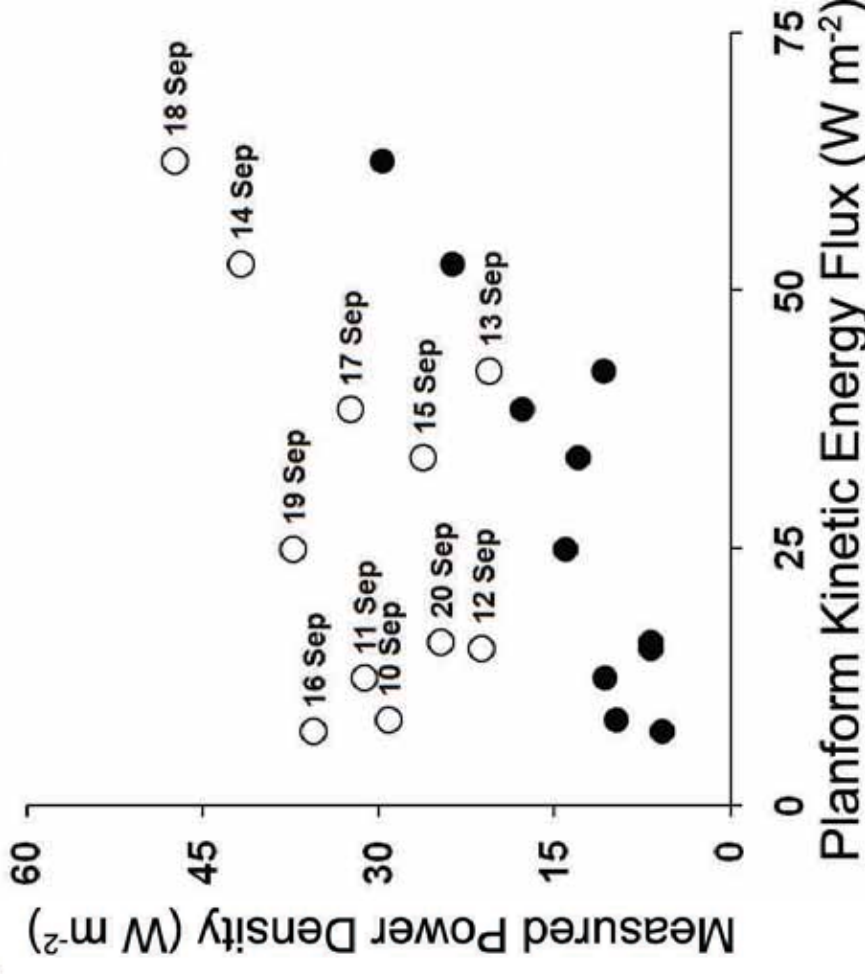

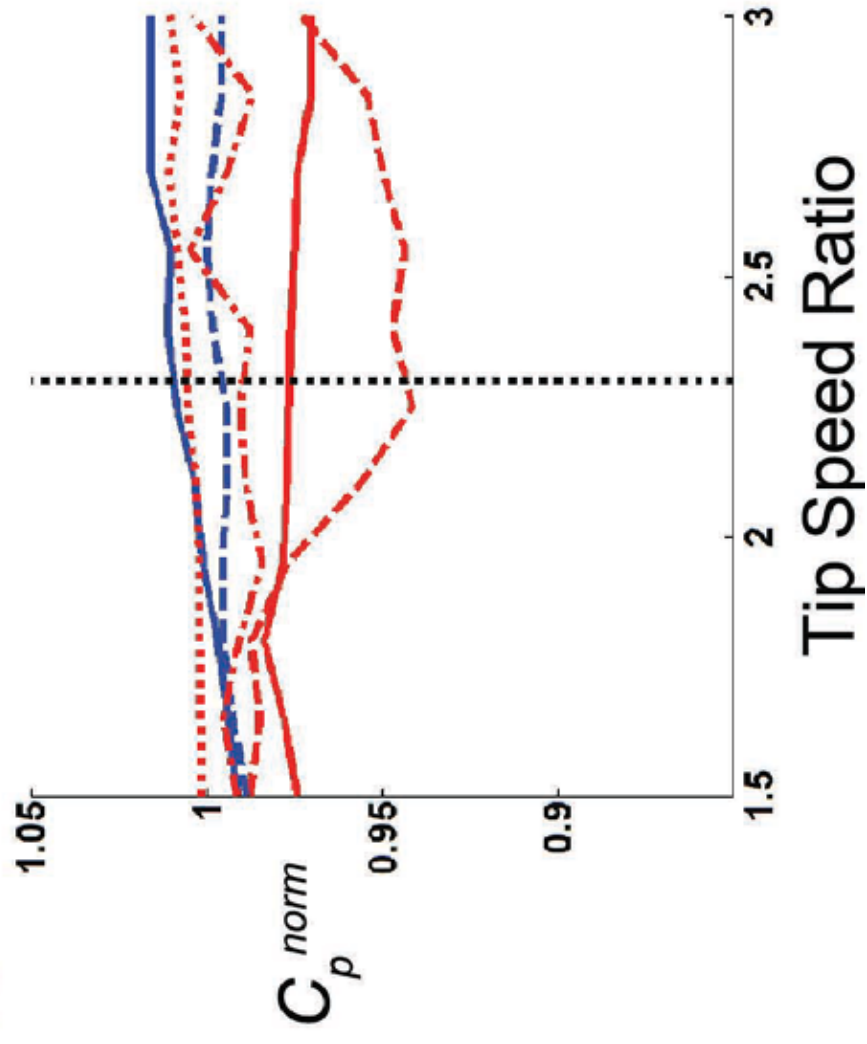

Figure 6

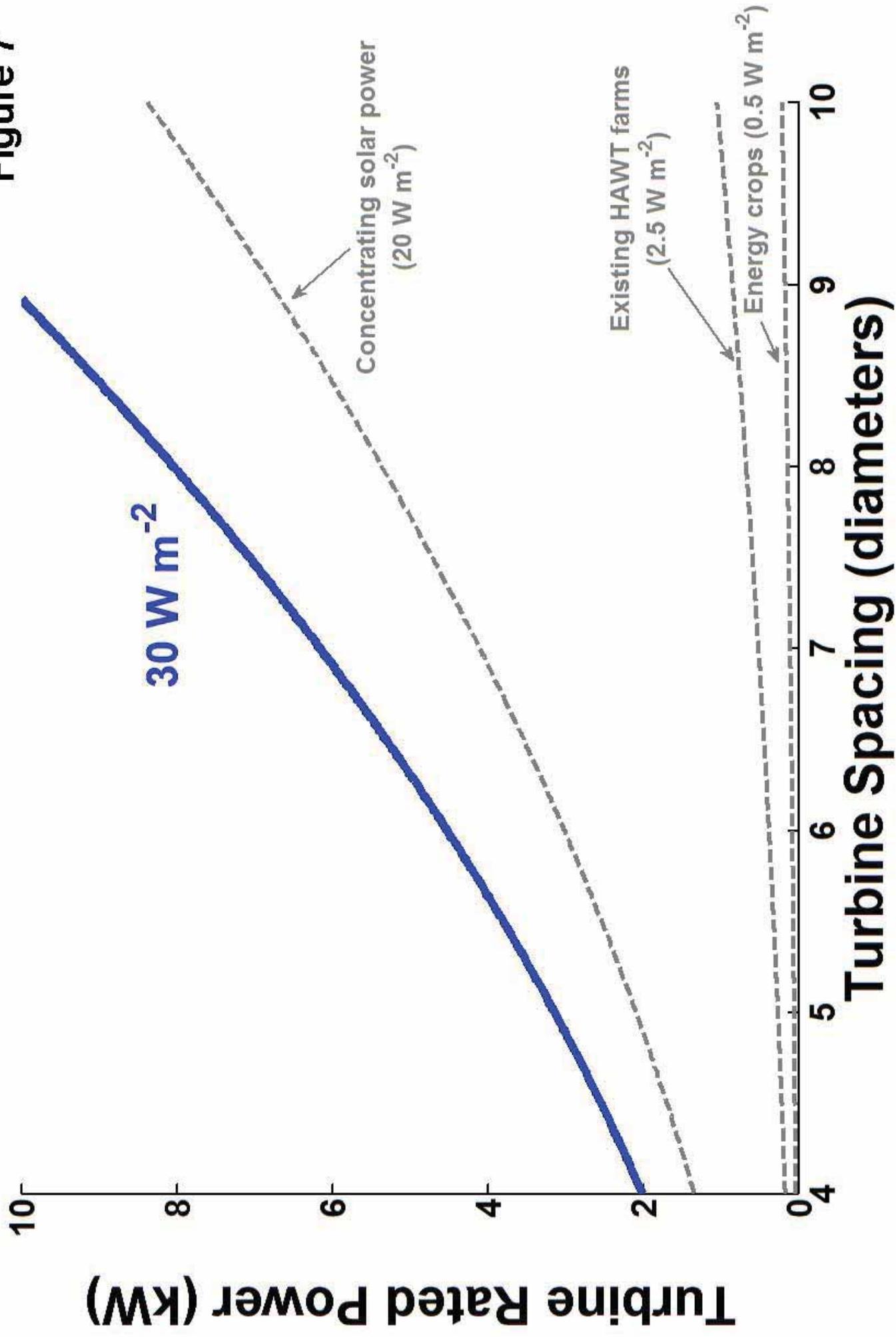

Figure 7

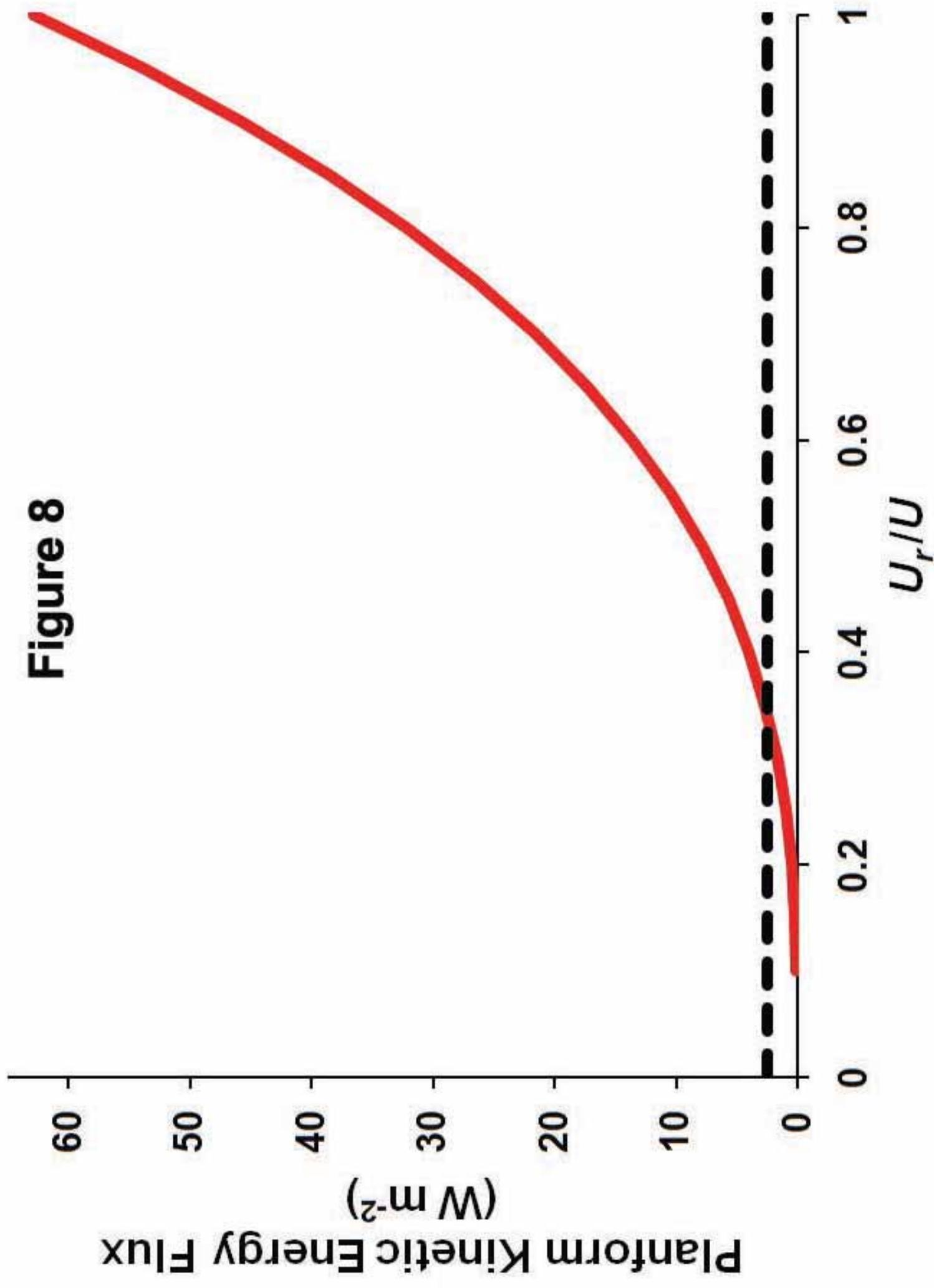

Figure 8

**Figure 9**

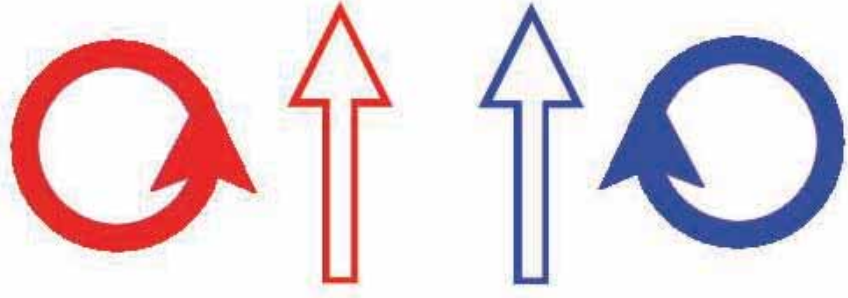

**B**

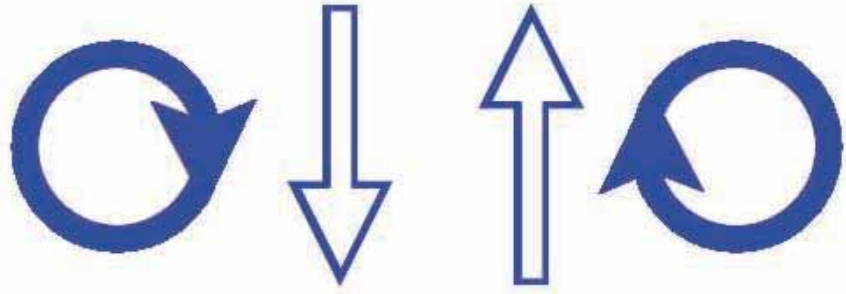

**A**

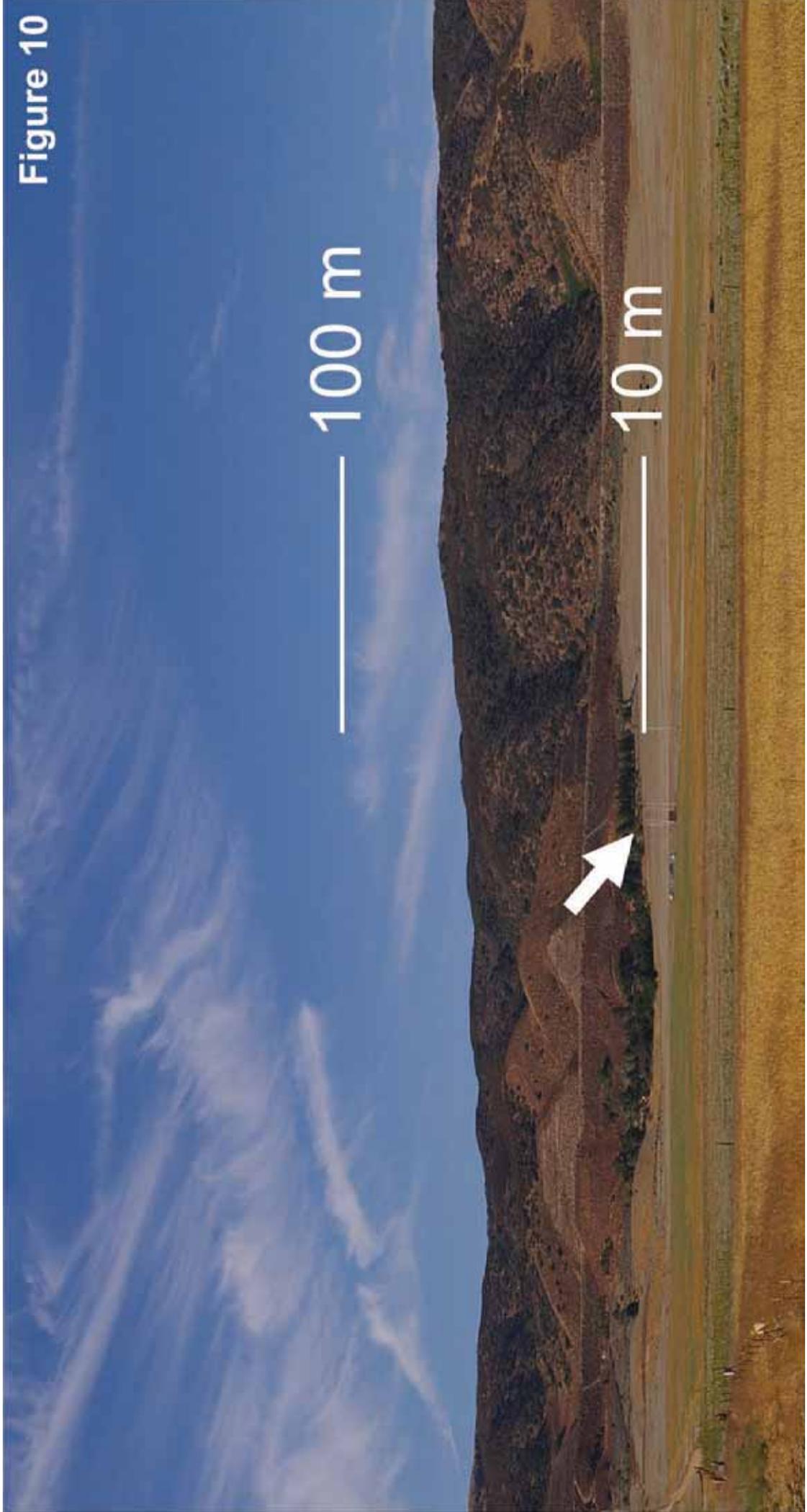

Figure 10

100 m

10 m